# Losing the battle over best-science guidance early in a crisis: Covid-19 and beyond

L. Illari[1], N. Johnson Restrepo[2], R. Leahy[1,2], N. Velásquez[3], Y. Lupu[4], N.F. Johnson[1*]

[1]Physics Department, George Washington University, Washington D.C. 20052
[2]ClustrX LLC, Washington D.C.
[3]Institute for Data, Democracy, and Politics, George Washington University, Washington D.C. 20052
[4]Department of Political Science, George Washington University, Washington D.C. 20052
*neiljohnson@gwu.edu

**Ensuring widespread public exposure to best-science guidance is crucial in a crisis, e.g. Covid-19, climate change. Mapping the emitter-receiver dynamics of Covid-19 guidance among 87 million Facebook users, we uncover a multi-sided battle over exposure that gets lost well before the pandemic's official announcement. By the time Covid-19 vaccines emerge, the mainstream majority -- including many parenting communities -- have moved even closer to more extreme communities. The hidden heterogeneity explains why Facebook's own promotion of best-science guidance also missed key audience segments. A simple mathematical model reproduces these exposure dynamics at the system level. Our findings can be used to tailor guidance at scale while accounting for individual diversity, and to predict tipping point behavior and system-level responses to interventions.**

Managing crises[1] such as the pandemic[2,3] and climate change[4] requires widespread public exposure to, and acceptance of, guidance based on best available science[5,6,7,8,9,10,11,12]. (Guidance is defined in the Oxford Dictionary[12] as "advice or information aimed at resolving a problem or difficulty"). But distrust of such best-science guidance has reached dangerous levels[7,8,9]. The American Physical Society, like many professional entities, is calling scientific misinformation one of the most important problems of our time[8,9]. During Covid-19's 2020 pre-vaccine period of maximal uncertainty and social distancing, many people went to their online communities for guidance about how to avoid catching it, and novel cures. Social media saw a huge jump[13] in users during 2020 (13.2%) taking the total to 4.20 billion (53.6% of the global population) with the top reason for going online given as seeking information[13].

Unfortunately, many people got exposed to guidance that was not best science[3] from their online communities of likely well-meaning but non-expert friends. Some even died as a result, e.g. drinking bleach or rejecting masks[14]. This raises the following urgent questions that we address here: Who emitted guidance to who? Who received guidance from who? What went wrong and when? And what does this tell us about how, where and when to intervene in current and future crises? Our answers follow from mapping the dynamical network of emitter-receiver guidance among online communities. This systems-level analysis hence complements -- but differs from -- the many excellent existing discussions about (mis)information[15,16,17,18,19,20,21,22,23,24,25,26,27,28,29,30,31,32,33,34,35,36,37,38,39,40,41,42,43,44,45,46,47,48,49,50,51,52,53,54,55].

Facebook is the dominant social media platform with 2.74 billion active users. Studies have shown that people (e.g. parents) rely on its in-built community structure for sharing guidance[56,57,58]. Hence we choose our main unit of analysis to be in-built Facebook communities -- specifically pages. We refer to each page simply as a community, but stress it is unrelated to any ad-hoc community structure inferred from network algorithms. Each page aggregates people around some common interest, is publicly visible and doesn't require us to access personal information. Our starting point is the ecology of such communities that were interlinked on Facebook around the vaccine health debate just prior to Covid-19 (November 2019, see Supplementary Material (SM)). A link from community (page) $i$ to community (page) $j$ exists when $i$ recommends $j$ to all its members at the page level ($i$ likes/fans $j$) as opposed to a page member simply mentioning another page: as a result, members of $i$ can at any time $t$ be automatically exposed to fresh content from $j$, i.e. $j$ emits and $i$ receives (see SM). Of course, all $i$'s members may not pay attention to it or believe it -- however, it has been shown experimentally and theoretically[59] that an online community can suddenly tip to an alternate stance in a reproducible way if it has a committed minority of just 25% with a certain belief[59].

The SM contains a list of communities (nodes) and links, together with replication code for the mathematical model and data analysis, on publication. Following the methodology of Ref. 60 (see SM Sec. 1 for a review) we obtained a list of 1356 interlinked communities (pages) from across countries and languages that we then classified into 3 main types: 211 'pro' communities (blue nodes, Fig. 1) comprising 13.0 million individuals, whose content actively promotes best-science health guidance (pro-vaccination); 501 'anti' communities (red nodes, Fig. 1) comprising 7.5 million individuals, whose content actively opposes this guidance (anti-vaccination); and 644 'neutral' communities (green nodes, Fig. 1(a)) comprising 66.2 million individuals, that had community-level links with pro/anti communities pre-Covid but whose content is focused on other topics such as parenting (e.g. child education), pets, organic food.

Reference 13 established empirically that a typical user only likes 1 Facebook page on average, so we can estimate the size of each community by its number of likes (fans). Page size typically ranges from a few hundred to a few million. Even if the actual size differs by some specific factor, our main conclusions remain valid since they depend on comparative numbers. The neutral communities are interlinked with the pros and antis, but have not expressed a specific stance on vaccines pre-Covid. We further categorize the neutral communities according to each's declared topic of interest. We find 12 categories, e.g. parenting, organic food-lovers, pet-lovers (see SM Sec. 2 for full discussion and examples). Further subdivision is possible but would lead to categories with too few communities.

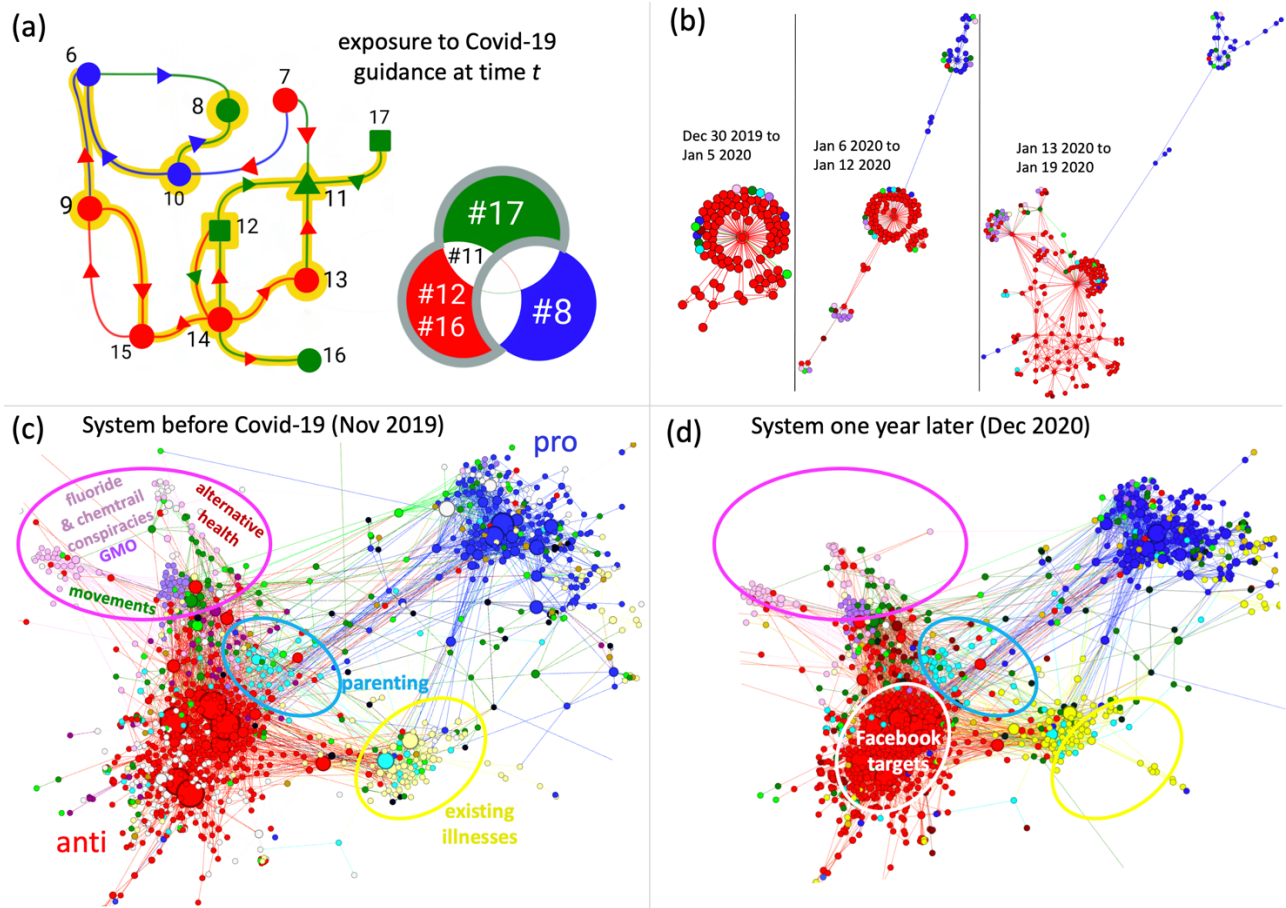

**Fig. 1: Exposure dynamics.** (a) Schematic illustrating emitter-receiver complexity. Each node is a community (Facebook page): pro communities (blue) actively promote best-science guidance, anti (red) actively oppose it. Neutrals (green) have a shape to denote their topic category (e.g. parenting). Link $i \to j$ means $i$ 'fans' $j$ which feeds content from page $j$ to page $i$, exposing $i$'s users to $j$'s content. Link $i \to j$ color is that of node $i$, arrow color is node $j$, arrow direction shows potential flow of Covid-19 guidance. Yellow indicates appearance of Covid-19 guidance at time $t$. Venn diagram shows source of neutral communities' exposure to Covid-19 guidance at time $t$. (b) Early evolution of exposure to Covid-19 guidance. Non-red/non-blue nodes in (b)-(d) denote categories of neutral communities, e.g. parenting communities are turquoise (SM Sec. 2 gives color scheme). Only links involving Covid-19 guidance during that time window are included (i.e. it is a filtered version of (a)). (c) Ecology pre-Covid, showing all potential links for exposure to Covid-19 guidance (unfiltered, as in (a)). Layout is spontaneous (ForceAtlas2) with closer proximity indicating more mutual links. Node size indicates community (i.e. page) size. (d) One year later, just before Covid-19 vaccine rollout. Nodes (pages) in white ring were the main targets of Facebook's banners promoting best-science guidance (see SM Sec. 3). Rings are in same position in (c) and (d) to highlight the increase in bonding.

Even with this oversimplified node and link classification, the exposure dynamics are highly complex: in Fig. 1(a), the color of a link from node (community, i.e. page) $i$ to node (community, i.e. page) $j$ is that of node $i$, the arrow color is that of node $j$, and the arrow's direction indicates potential flow of Covid-19 guidance. If a node $j$ posts Covid-19 guidance at time $t$, we put a yellow border around it and around any arrows emanating from it, to indicate exposure of the linked nodes to $j$'s Covid-19 guidance at time $t$. If there are no links going into node $j$, it is not exposing any other node to its Covid-19 guidance. Each page (node) could link to various other pages, but irrelevant links get filtered out as explained in Ref. 60 and SM Sec. 1 yielding a network with a few links per node. The Venn diagram shows the Covid-19 guidance exposure of the neutral communities at time $t$ (green nodes, each shape is a separate topic): the gray border contains the nodes exposed to Covid-19 guidance that comes entirely from non-pro communities, i.e. it comes from antis and/or other neutrals. For example, 12 is only exposed to Covid-19 guidance from anti node 14, hence 12 is in the anti-only gray-border region. 12 has a link into 11 and so 11 is exposed to Covid-19 guidance from 12, but this doesn't affect 12 itself.

Given the complexity already in Fig. 1(a), we need to simplify other aspects of this paper's analysis. We will not determine an absolute fraction of scientific truth in each piece of Covid-19 guidance: this is in any case challenging since even the wildest anti content can contain truthful fragments, e.g. Ref. 61 shows experimentally that quantum dots can act as injectable vaccine markers[62]. Instead, manual analysis of all the community content confirms the expectation that pro communities do indeed promote best-science Covid-19 guidance, e.g. from Centers for Disease Control (CDC); anti communities' guidance opposes this; and neutral communities' guidance sits in between getting further downgraded by additional non-scientific comments posted by page members. Even if a significant fraction of our classifications is wrong, the main conclusions are unchanged since they only depend on relative numbers: we checked this explicitly by introducing a 15% error into our classifications. We recognize that best-science guidance can change over time, and may eventually be proven wrong -- but that is rare.

Figure 1(b) shows how the struggle over Covid-19 guidance develops well before the official announcement of the pandemic on March 11. It is a filtered version of the construction in Fig. 1(a), i.e. a link only appears when one of the nodes (communities) that it connects presents Covid-19 guidance in that time interval. It shows the giant connected component. Because we use the ForceAtlas2 layout algorithm, the observed segregation is self-organized and proximity indicates stronger mutual links, i.e. the more links node $i$ and its neighbors have with node $j$ and its neighbors, the closer visually node $i$ will be to node $j$ (see SM Sec. 4). It reveals how quickly anti communities (red nodes) dominate, with neutrals (not red nor darker blue, e.g. parenting communities are turquoise) also getting picked up or attaching themselves. Pro communities (darker blue) appear later, in a separated structure. This pro-anti segregation suggests that the observed system strengthening from Fig. 1(c) to 1(d) derives from the early bonding around Covid-19 guidance.

Figure 1(d) provides a system-level view just before Covid-19 vaccine rollout, of all the links along which Covid-19 guidance could flow (it is equivalent to Fig. 1(a) ignoring the yellow shading, and hence akin to a road network irrespective of the traffic while Fig. 1(b) is the subset of roads carrying traffic). Since proximity in the ForceAtlas2 layout indicates stronger mutual links (see SM Sec. 4), the closer nodes appear spatially then the more likely they are to share content. The observable changes from Fig. 1(c) indicate that not only did the antis tighten internally during this period of maximal societal uncertainty prior to vaccines appearing, the neutrals were pulled and/or pulled themselves closer to the antis, and neutral categories such as the parenting communities (turquoise nodes) also tightened internally.

This has the key consequence that by the time buy-in to these new vaccines was becoming essential (i.e. December 2020), many parents who were responsible for health decisions about themselves, their young children and also likely elderly relatives, had become even closer to the antis who held extreme views including distrust of vaccines and rejection of masks -- and also to other neutrals (see nodes within purple ring) who were focused on non-vaccine and non-Covid-19 conspiracy content surrounding climate change, 5G, fluoride, chemtrails, GMO foods, and also alternative health communities that believe in natural cures for all illnesses (see SM Sec. 2 for details).

Facebook conducted its own top-down promotion of best-science Covid-19 guidance by placing banners at the top of some pages (i.e. nodes) pointing to the CDC for example (see SM Sec. 3). However, these banners appear primarily in anti communities (red nodes) and the antis that were targeted were primarily within the white oval in Fig. 1(d) (SM Sec. 3). Hence most neutral communities were missed -- yet this would have been avoidable using these exposure maps.

The Venn diagram in Fig. 2(a) quantifies the extent to which non-pro communities acted as the dominant sources (i.e. emitters) of Covid-19 guidance to neutral communities during the period of maximal societal uncertainty prior to vaccine discovery. While 7.19 million individuals were exposed exclusively to Covid-19 guidance from non-pro communities, only 1.28 million were exposed exclusively to Covid-19 guidance from pro communities. The remaining 5.40 million were exposed to both, which may still have made them quite uncertain about what to think. Figure 2(b) shows this imbalance was even worse for individuals in parenting communities: 1.10 million of these individuals were exposed exclusively to Covid-19 guidance from non-pro communities, but only 503 were exposed exclusively to Covid-19 guidance from pro communities.

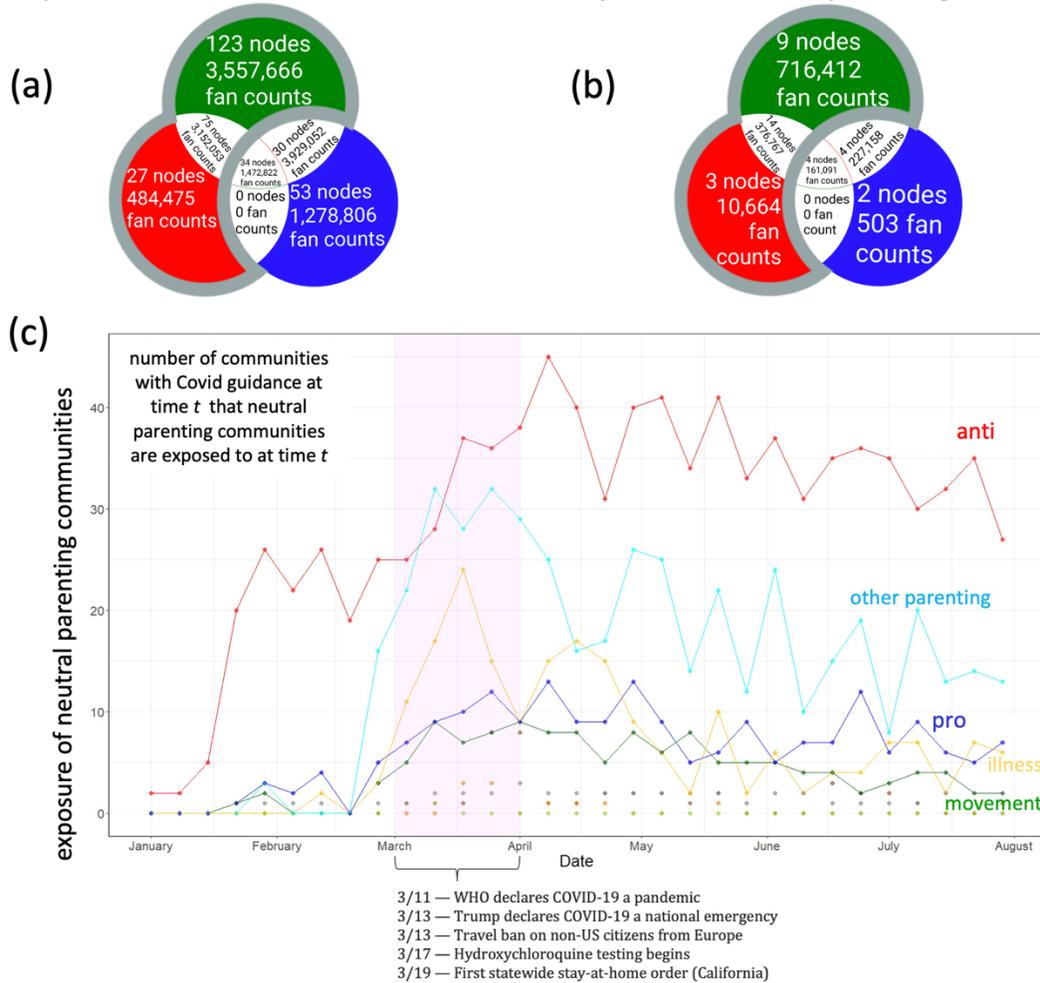

**Fig. 2:** (a) Venn diagram, as in Fig. 1(a), shows sources of exposure to Covid-19 guidance for all neutral communities in the giant connected component of our ecology (Fig. 1(c)(d)) aggregated over Jan-August 2020. (b) Similar to (a) but just for the parenting community subset of all neutral communities. (c) Parenting communities' exposure to Covid-19 guidance disaggregated over time and by source (i.e. emitter) type.

Disaggregating this further, Fig. 2(c) shows the sources of parenting communities' exposure to Covid-19 guidance over time. Starting in early January, the anti communities quickly generated Covid-19 guidance which, when combined with the significant number of links with parenting communities in Fig. 1(c), generated the rapid rise in parenting communities' exposure from anti communities shown in Fig. 2(c). Instead of pro communities then stepping in, there was next a rapid rise of exposure to guidance from other parenting communities -- all well before the official declaration of a pandemic. This was accompanied by a smaller rise in exposure from communities focused around pre-existing, non-Covid illness such as Asperger's Syndrome and cancer. These high levels of exposure from anti and other parenting communities persisted for the entire period. In contrast, exposure from the pro communities never showed any strong response and remained surprisingly low. The SM Sec. 5 shows that these curves in Fig. 2(c) are statistically significant as compared to a null model in which a random network is chosen, meaning we can reject the hypothesis that the microstructure of the exposure network is not relevant. In short, the complexity of the exposure network (Fig. 1) is key to understanding the exposure dynamics over time.

These findings paint the following picture of the pre-vaccine period: individuals in the neutral parenting and other mainstream communities became aware of Covid-19 guidance from anti communities early in January 2020, which they then quietly deliberated over, perhaps interacting in private groups or apps such as WhatsApp or with others offline. By mid-February, they felt in a position to produce and share their own Covid-19 guidance with communities like theirs (parenting). Meanwhile, they only received minimal best-science guidance from the pro communities (dark blue curve is near zero). They did not have any strong tendency to create new links to other pro communities -- probably because they were instead receiving guidance from other neutral communities that have similar interests (e.g. parenting) who they feel they can identify with, and hence perhaps trust more.

A broader implication is that actionable tipping points can arise even before any official announcement of a crisis. While direct messaging against the antis during their January rise as guidance emitters (red curve, Fig. 2(c)) may not have been desirable given their active opposition and possible backlash, the February rise of other parenting communities as guidance emitters (turquoise curve) was a missed opportunity for outreach. Best-science Covid-19 guidance from the pros could instead have been tailored around the popular topics within the parenting communities at that time, which could have been read from their pages, and hence introduced at scale using the map in Fig. 1(b). Even a blanket intervention across all neutral categories using Fig. 1(c), would likely have reduced the subsequent high peaks in exposure to Covid-19 guidance from non-pro communities and its persistence (Fig. 2(c)).

This raises the need for a mathematical understanding of the science of such online tipping points and potential interventions. We derive our multiscale mathematical model in the SM Sec. 6, built around the fact that individuals 'gel' together online and these gels (pages, or clusters of pages) may influence each other through their content.

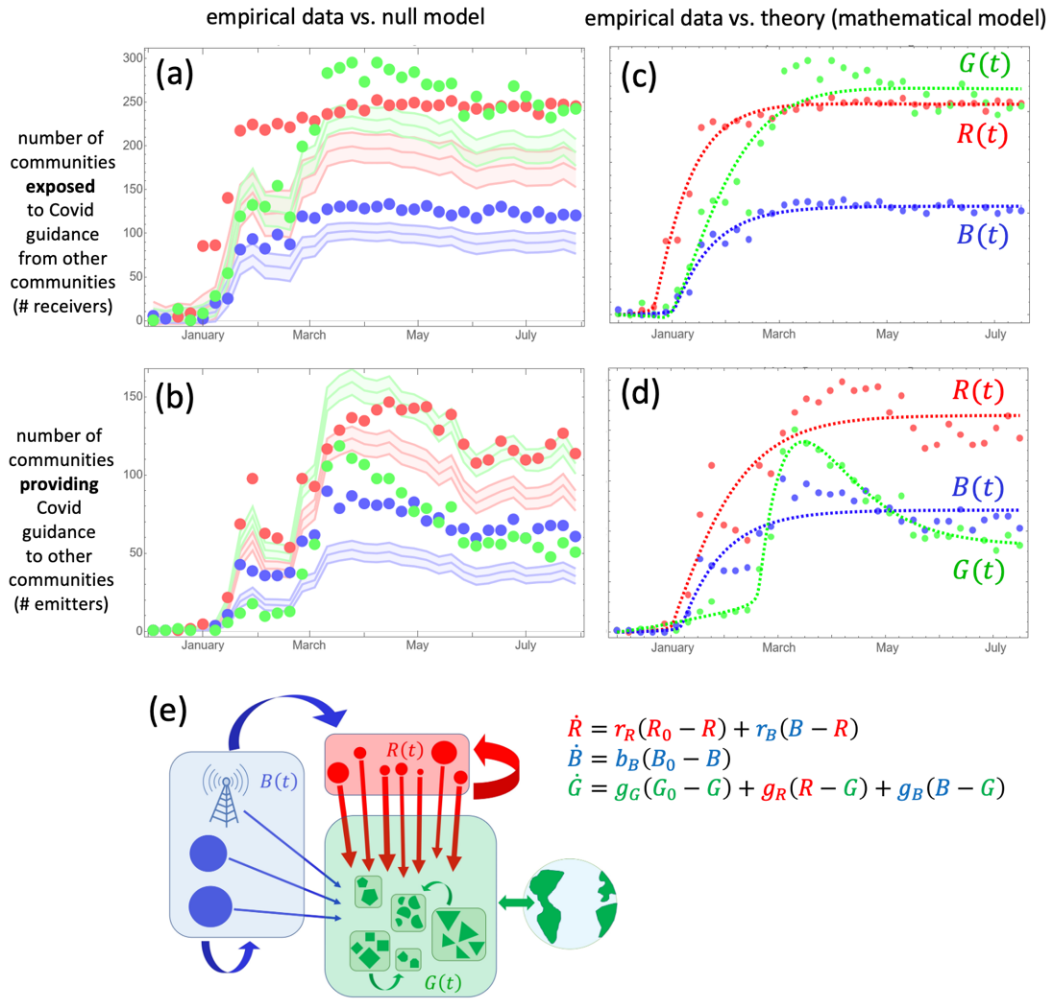

**Fig. 3:** (a) Empirical data (circles) shows number of pro (blue), anti (red), neutral (green) communities exposed to Covid-19 guidance (i.e. receivers). Lines show range of outputs from the null model, which provides a poor fit to the data. (b) Similar to (a) but circles show number of communities providing Covid-19 guidance (i.e. emitters). (c) and (d) compare the empirical data to our generative mathematical model (dotted lines) using two sets of coupling values in (e) (see SM Sec. 6 and software files for replication). (e) Simplest version of our model (SM Sec. 6) in which pros, antis and neutrals are aggregated over all pages (shapes) with neutrals also aggregated over all 12 categories (boxes).

Figure 3(e) shows a minimal linear version at the scale of the aggregated pros (blue, $B(t)$), antis (red, $R(t)$) and neutrals (green, $G(t)$). It is consistent with the empirical observations that (i) neutrals can be impacted by guidance from antis, pros and other neutrals including the rest of the mainstream global population, hence the couplings of $\dot{G}(t)$ to $R(t)$ with strength $g_R$, to $B(t)$ with strength $g_B$, and to $G(t)$ with strength $g_G$; (ii) antis can be impacted by guidance from other antis, and by pros whose guidance they ridicule openly on their pages, hence the couplings of $\dot{R}(t)$ to $R(t)$ with strength $r_R$, and to $B(t)$ with strength $r_B$; (iii) pros promote best-science guidance which only depends on the scientific advances that other pros may be reporting, hence $\dot{B}(t)$ only couples to $B(t)$ with strength $b_B$. This simple version (Fig. 3(e)) gives visually close agreement with the data, and this agreement gets even better using a version that incorporates the finite gel onset times predicted by the full mathematical theory (see SM Sec. 6 and Figs. 3(c)(d)). The aggregate exposure of the neutral communities ($G(t)$ in Figs. 3(a)(c)) rises to a persistent peak that exceeds $R(t)$ and $B(t)$, confirming the importance of the exposure dynamics among the neutrals. We build a null model by shuffling node identities separately within the antis, the pros and the 12 neutral subcategories, hence maintaining the numbers in each subcategory but losing their locations in the network. Repeating this 1000 times yields the bands in Figs. 3(a)(b) which are far from the empirical data (see SM Mathematica and data replication files for full details), hence the importance of the actual links and the full network in determining the online exposure dynamics.

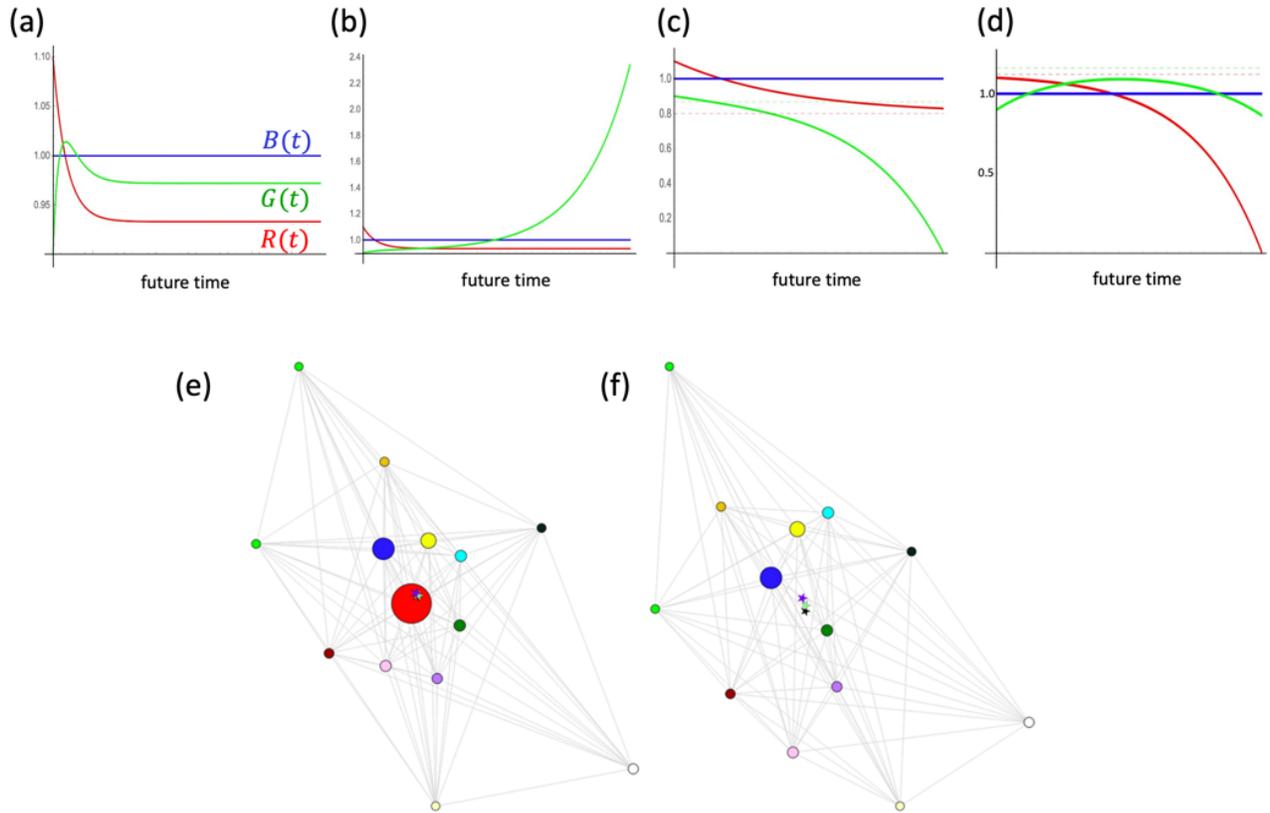

**Fig. 4:** (a)-(d) The four classes of future outcome predicted by the mathematical model (Fig. 3(e)) starting with initial conditions that crudely mimic the current situation (SM Sec. 6 Figs. S13-15 show details and code). (e) shows a renormalized version of Fig. 1(d) in which nodes of a given type are aggregated into a single super-node with that weighted size/mass. The center of this online universe is shown using various definitions: spatial center (black star), center weighted by degree (purple star), center weighted by number of clusters (green star). (f) shows the impact on (e) of removing the anti (red) super-node: the pro (blue) super-node still does not sit at the center of the new universe.

To our knowledge, this represents the first generative mathematical model of online exposure to guidance that also accounts for the reality that an online audience comprises interconnected communities. It can help address questions about (mis)information and harms in the online world, as well as making predictions and exploring what-if interventions. Figures 4(a)-(d) illustrate this using a crude estimate of current $R(t)$, $B(t)$ and $G(t)$ values, as shown in SM Figs. S13 and 14. Based on the current persistence of hesitancy about vaccines and mask wearing, we take the pros as having reached their maximum capability in terms of promoting best-science guidance, hence $B(t)$ is constant. It predicts four classes of future (see SM Sec. 6 Figs. S13-15 for proof). Figure 4(a) shows what happens when the future couplings between anti, pro and neutral are all positive (i.e. positive feedback): $G(t)$ initially peaks before settling at a higher value. In (b), neutral and pro have negative coupling (i.e. negative feedback): this causes $G(t)$ to escalate dramatically. In (c), all couplings are negative: $G(t)$ drops dramatically. In (d), neutral and anti have the only positive coupling: $R(t) \to 0$ but $G(t)$ remains high for an extended period of time.

These mathematical predictions can then be tied in with the network picture from Figs. 1-2, using the physics technique of renormalization in which the communities of the anti, pro and the 12 neutral subcategories are each aggregated into their own community-of-communities ball. For example, Fig. 4(f) shows the impact of removing the antis (c.f. Fig. 4(e)) hence mimicking Fig. 4(d) in which $R(t) \to 0$. The comparison is only valid at short times since we are not allowing the network as a whole to adapt or rewire after cutting all anti links. Figure 4(f) shows that the pro communities will still not sit at the center of this online universe because of the many-sided interactions with the neutral subcategory communities, particularly the movement communities (dark green ball). The fact that both neutrals and pros remain in play in Fig. 4(f) is broadly consistent with Fig. 4(d) in which $G(t)$ remains high and comparable to $B(t)$ for an extended period despite $R(t) \to 0$.

**Supplementary Material (SM)**
A list of communities (nodes) and links, together with full computer (Mathematica, Excel) files for replication of the mathematical analysis and data fitting in this study, will be provided on publication.
**Section 1:** Methodology for collecting data, building network, and analysis in the paper
**Section 2:** Color scheme for neutral nodes in plots. Classification of neutral nodes
**Section 3:** Example of Facebook banners promoting best-science Covid-19 guidance. Positions in network of the nodes that receive Facebook banners promoting best-science Covid-19 guidance
**Section 4:** ForceAtlas2 layout and analysis showing dependence of layout on strength of bonding
**Section 5:** Statistics of the fit in Fig. 3(c)(d). Null model analysis throughout the paper
**Section 6:** Derivation of our mathematical model and its use in Fig. 3, and predictions from our mathematical model shown in Fig. 4(a)-(d)

This material is based upon work supported by the Air Force Office of Scientific Research under award numbers FA9550-20-1-0382 and FA9550-20-1-0383. Any opinions, findings, and conclusions or recommendations expressed in this material are those of the author(s) and do not necessarily reflect the views of the United States Air Force.